\documentclass[a4paper]{article}

\usepackage{INTERSPEECH2020}

\usepackage{color}
\usepackage{diagbox}
\usepackage{xcolor,colortbl}
\usepackage{enumitem}
\usepackage{amssymb}
\usepackage{balance}

\newcommand{\eg}{e.\,g.\,, }

\usepackage[activate]{microtype}
\usepackage{times}
\usepackage{graphicx}
\usepackage{amssymb}
\usepackage{gensymb}
\usepackage{amsmath}
\usepackage{breakurl}
\usepackage{caption}
\usepackage{subcaption}
\usepackage{tikz}
\usetikzlibrary{shapes.geometric, positioning, arrows, bayesnet}

\usepackage{url,hyperref}

\usepackage{algorithm}
\usepackage{algorithmic}

\usepackage[labelfont={bf},font=small]{caption}
\usepackage[none]{hyphenat}

\usepackage{mathtools, cuted}

\usepackage[noadjust, nobreak]{cite}

\usepackage{tabularx}
\usepackage{amsmath}

\usepackage{float}

\usepackage{pifont}

\newcolumntype{Y}{>{\centering\arraybackslash}X}

\usepackage[]{placeins}

\usepackage{placeins}

\usepackage{tikz}

\usepackage[framemethod=tikz]{mdframed}

\usepackage{afterpage}

\usepackage{stfloats}

\usepackage{atbegshi}
\newcommand{\handlethispage}{}
\newcommand{\discardpagesfromhere}{\let\handlethispage\AtBeginShipoutDiscard}
\newcommand{\keeppagesfromhere}{\let\handlethispage\relax}
\AtBeginShipout{\handlethispage}

\usepackage{comment}

\title{ConcealNet: An End-to-end Neural Network for Packet Loss Concealment in Deep Speech Emotion Recognition}

\name{Mostafa M.\ Mohamed$^{1,2}$ and Bj\"orn W.\ Schuller$^{2,3}$}
\address{
  $^1$Chair of Embedded Intelligence for Health Care and Wellbeing, University of Augsburg, Germany\\
  $^2$AI R\&D team, SyncPilot GmbH, Augsburg, Germany\\
  $^3$GLAM -- Group on Language, Audio,   \& Music, Imperial College London, UK}
\email{mostafa.amin@syncpilot.com, schuller@IEEE.org}

\begin{document}

\maketitle
\thispagestyle{empty}
\pagestyle{empty}

\begin{abstract}
Packet loss is a common problem in data transmission, including speech data transmission. This may affect a wide range of applications that stream audio data, like streaming applications or speech emotion recognition (SER). Packet Loss Concealment (PLC) is any technique of facing packet loss. Simple PLC baselines are 0-substitution or linear interpolation. In this paper, we present a concealment wrapper, which can be used with stacked recurrent neural cells. The concealment cell can provide a recurrent neural network (ConcealNet), that performs real-time step-wise end-to-end PLC at inference time. Additionally, extending this with an end-to-end emotion prediction neural network provides a network that performs SER from audio with lost frames, end-to-end. The proposed model is compared against the fore-mentioned baselines. Additionally, a bidirectional variant with better performance is utilised.
For evaluation, we chose the public RECOLA dataset given its long audio tracks with continuous emotion labels. ConcealNet is evaluated on the reconstruction of the audio and the quality of corresponding emotions predicted after that. The proposed ConcealNet model has shown considerable improvement, for both audio reconstruction and the corresponding emotion prediction, in environments that do not have losses with long duration, even when the losses occur frequently.
\end{abstract}


\section{Introduction}
Packet Loss Concealment (PLC) is any technique that attempts to handle the effects of packet loss or overly delayed packets. This is a common problem in speech transmission using VoIP \cite{voip}. This problem can affect the performance of many speech processing systems that assume a complete speech signal is transmitted, including Speech Emotions Recognition (SER).
There has been a variety of classical techniques that attempts to solve the packet loss problem, for example using Hidden Markov Models (HMM) \cite{hmminterspeech} and Linear Predictive Coding (LPC) \cite{schuller2013}. There are also encoding-based techniques \cite{codec}. However, in the era of deep learning and the rise of a variety of generative networks, like sequential generative Recurrent Neural Networks \cite{graves} and Generative Adversarial Networks (GAN) \cite{GANS}, generating data in place of the lost packets is a promising avenue for advanced concealment techniques.
There exist studies \cite{Lotfidereshgi2018} that attempt to solve packet loss in the context of Automatic Speech Recognition (ASR). However, to the authors' best knowledge, there are no studies addressing PLC in the context of Speech Emotion Recognition.
{In an earlier work, the authors have investigated techniques how to train SER end-to-end models to be robust in the presence of frame-loss \cite{mywork}. However, we attempt here to address PLC directly to address this issue.} 
Furthermore, this problem can happen on a variety of devices including mobile devices \cite{Marchi16-RTO}. Providing a neural network that can perform PLC end-to-end would be favourable, because it is easier to embed it into different applications without the need for extra processing. More importantly, there is a rise nowadays of hardware optimised for neural networks processing \cite{smartphones}, hence having an end-to-end PLC solution would be the most suitable solution for future hardware. 
The contributions of this paper are providing such an end-to-end PLC neural network (we call it ConcealNet) and examining the effects of using it on SER in lossy environments.
    

The paper is divided as follows: in Section \ref{sec:related}, we will review some existing techniques that are also used for PLC with different models or with different settings. 
In Section \ref{sec:approach}, the main approach will be presented. The experiments and evaluations are discussed in Section \ref{sec:results}. Finally, in Section \ref{sec:conclusion}, the summary and the conclusion of the paper are discussed.

\section{Related work}
\label{sec:related}

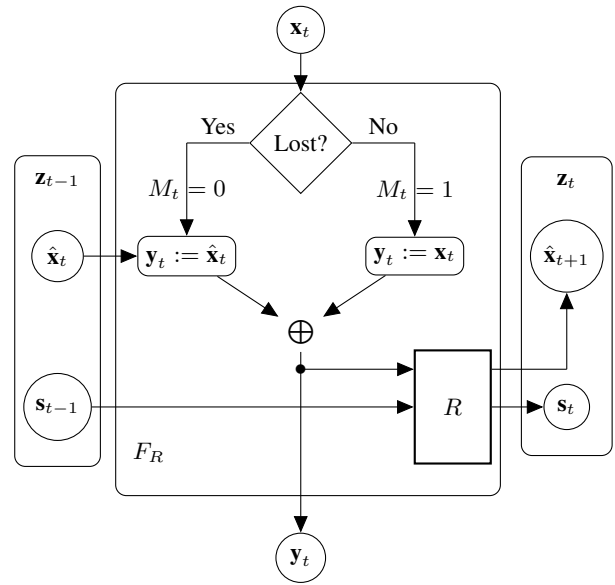
\begin{figure}[h!]
    \centering
    \begin{tikzpicture}
        \node[circle, draw] at (0, 0) (qn) {$\textbf{y}_t$};

        \node[circle, draw] at (3.5, 2) (St)   {$\textbf{s}_t$};
        \node[circle, draw] at (3.5, 4) (Xt1) {$\hat{\textbf{x}}_{t+1}$};
        \node[circle, draw] at (-3.2, 4) (xp) {$\hat{\textbf{x}}_{t}$};
        \node[circle, draw] at (-3.2, 2) (sp) {$\textbf{s}_{t-1}$};
        
        \node [rectangle] at (2, 2) (R) [draw,thick,minimum width=1cm,minimum height=1.5cm] {$R$};      
        \node [rectangle, draw, minimum height=0.5cm, rounded corners]  at (1.5, 4) (yex) {$\textbf{y}_t := \textbf{x}_t$};
        \node [rectangle, draw, minimum height=0.5cm, rounded corners]  at (-1.5, 4) (yexp) {$\textbf{y}_t := \hat{\textbf{x}}_t$};

        \node [diamond, draw] at (0, 5.5) (lost) {Lost?};
        \node[circle, draw] at (0, 7) (inpx) {$\textbf{x}_t$};
        \node [draw=none,fill=none] at (0, 3) (conc) {$\bigoplus$};
        \node [draw=none,fill=none] at (2.4, 2.5) (aux2) {};
        \node [draw=none,fill=none] at (0, 2.5) (aux3) {$\bullet$};
        \node [draw=none,fill=none] at (1.6, 2.5) (aux4) {};

        \node [] (FR) at (-2, 1.4) {$F_R$};
        \node [] (zp) at (-3.2, 5) {$\textbf{z}_{t-1}$};
        \node [] (zt) at (3.5, 5) {$\textbf{z}_t$};
        \plate[] {plate1} {(zp)(xp)(sp)} { };
        \plate[] {plate2} {(St)(Xt1)(zt)} { };
        
        \plate[] {plate3} {(R)(lost)(yex)(yexp)(FR)} { };
        \draw (R) [->] edge (St);
        \draw [-] (aux2) -- (3.5, 2.5);
        \draw [->] (3.5, 2.5) --  (Xt1);
        \draw [->] (0, 2.5)  -- (aux4);
        \draw (sp) [->] edge (R);
        
        \draw (lost) [-, above] edge node{No} (1.5, 5.5);
        \draw [->] (1.5, 5.5) -- node{$M_t=1$} (yex);
        
        \draw (lost) [-, above] edge node{Yes} (-1.5, 5.5);
        \draw  [->] (-1.5, 5.5) -- node{$M_t=0$} (yexp);
        
        \draw (inpx) [->] edge (lost);
        \draw (xp) [->] edge  (yexp);
        \draw (yexp) [->] edge (conc);
        \draw (yex) [->] edge  (conc);
        \draw (conc) [->] edge (qn);
    \end{tikzpicture}

    \caption{Demonstration of how the recurrent concealment cell $F_R$ operates. The input is a frame $\textbf{x}_t$ with a corresponding mask element $M_t$. $\textbf{x}_t$ is lost if $M_t=0$, otherwise $M_t=1$.
    $\textbf{y}_t$ is then calculated, which is the predicted original (or concealed in case of loss). Then the wrapped recurrent cell $R$
    is used to predict the next frame $\hat{\textbf{x}}_{t+1}$. $R$ is a (stacked) recurrent cell.}
    \vspace{-0.35cm}
        \label{fig:concealment}
\end{figure}

A PLC approach is proposed in \cite{PLC2016}, which relies on features representing speech data. The concealment is done on feature level and then decoded, rather than executing it on the actual speech directly. The approach was realised in the context of enhancing Automatic Speech Recognition (ASR). Based on \cite{PLC2016}, \cite{Lotfidereshgi2018} implemented a PLC algorithm that operates directly on the speech data using LSTM-based neural networks. They also applied it on ASR, while evaluating on the TIMIT dataset \cite{TIMIT}. The main advantage of both approaches is that, they can be applied in a frame-by-frame fashion, which is suitable for real-time application on losses of small packets. Additionally, they have the potential to be extended to a neural-based end-to-end PLC.
GAN-based approaches are utilised in \cite{GAN, acousticinpainting}, where the generator adapts an architecture similar to an auto-encoder. The model uses audio of long segments (like 3\,200\,ms) to make predictions, which is longer than a typical packet size in VoIP being around 10-20\,ms \cite{voip, PLC2016}. Such a setup is most effective for offline processing and for long losses.  
%
In \cite{xiao2018packet}, there is an approach facing PLC not in the context of speech, but rather the transmitted data from pose tracking sensors. They authors also chose LSTM-based RNNs, while having a two-state Markov Chain for packet loss injection.
\cite{khan2018} are considering Cartesian Genetic Programming \cite{Miller2011} for signal reconstruction, in some abstract setting.
Another approach like \cite{mack2019deep} attempts to reconstruct STFT signals, under a variety of deformations like destructive interference and packet-loss, however, it is not directly addressing PLC in particular.



\section{Approach}
\label{sec:approach}

\subsection{Recurrent generative modelling}
\label{subsec:generativeRNN}

Given an input sequence $\textbf{x}_1, \cdots , \textbf{x}_T$, where $\textbf{s}_{t-1}$ denotes the previous state of $\textbf{x}_t$ representing the whole preceding sequence, then, a recurrent neural cell is an operation $R$ that computes an output $\textbf{y}_t$ and a next state $\textbf{s}_t$ \cite{deeplearn}. Two effective and commonly used recurrent cells are gated recurrent cells like LSTM \cite{LSTM} and GRU \cite{GRU}. Additionally, $R$ here could also refer to a stack of recurrent cells, such that at each time step, each cell takes input from the output of the preceding cell at the same time step.

\begin{table}[b!]
    \centering
    \vspace{-0.5cm}
    \begin{tabular}{c|c|c}
        Operation & specs & output shape  \\
        \hline
        Input & & $(T, 100)$ \\
        \hline
        LSTM & 768 & $(T, 768)$ \\
        LSTM & 768 & $(T, 768)$ \\
        \hline
        Fully Connected & 256 & $(T, 256)$ \\
        Fully Connected & 100 & $(T, 100)$ \\
    \end{tabular}
    \vspace{0.2cm}
    \caption{Generative RNN architecture that predicts a frame using the preceding frames.}
    \vspace{-0.5cm}
    \label{tab:predictive_coding}
\end{table}

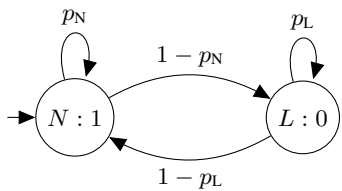
\begin{figure}[b!]
    \centering
    \begin{tikzpicture}
        
        \node[circle, draw] (qn) {$N:1$};
        \node[] at (-1, 0) (start) {};
        \node[circle, draw] at (3, 0) (ql) {$L:0$};
        
        \draw 
        (start) [->] edge (qn)
        (qn) edge[loop above] node{$p_{\text{N}}$} (qn)
        (ql)[->] edge[loop above] node{$p_{\text{L}}$} (ql)
        (qn)[->] edge[bend left, above] node{$1 - p_{\text{N}}$} (ql)
        (ql) edge[bend left, below] node{$1 - p_{\text{L}}$} (qn);
    \end{tikzpicture}

    \caption{Markov Chain $\mathcal{M}(p_{\text{L}}, p_{\text{N}})$ that samples a binary sequence, that can be used as a mask for {\it loss} or {\it non-loss}.}
        \label{fig:markov}
\end{figure}
As shown in \cite{graves}, generative Recurrent Neural Networks (RNNs) can be used to generate data by training the cells to predict elements of sequences using the preceding elements. Inspired by this, we train a similar approach as a regression task instead of classification, to enable an RNN $G$ to generate audio segments. $G$ will be later used to conceal packet loss by generating audio segments for the lost packets. 
\\
\\
\textbf{Training.}
We train the generative RNN $G$, using the input speech segments $\textbf{x}_1, \textbf{x}_2, \cdots, \textbf{x}_{T - 2}, \textbf{x}_{T-1}$ and the predicted output ($\textbf{y} = G(\textbf{x})$) which is a sequence $\textbf{y}_1, \textbf{y}_2,  \cdots,\textbf{y}_{T - 2}, \textbf{y}_{T-1}$, by concatenating the segments and comparing that to the concatenation of the sequence $\textbf{x}_2, \cdots, \textbf{x}_{T - 1}, \textbf{x}_{T}$ as the corresponding ground truth prediction. This is optimising the loss function $\mathcal{L}(\textbf{x}_{2 \cdots T}, \textbf{y}_{1 \cdots T  -1 })$. The loss is $1 - \rho_c(x, y)$, where $\rho_c(x, y)$ is the \textit{concordance correlation coefficient} (CCC) which measures data reproducibility \cite{ccc}, given by:
\begin{equation}
\rho_c(x, y) = \frac{2\sigma_{xy}^2}{\sigma_x^2 + \sigma_y^2 + (\mu_x - \mu_y)^2},
\end{equation}

where $\mu_x, \mu_y$ are the means, $\sigma_x^2, \sigma_y^2$ are the variances, and $\sigma^2_{xy}$ is the covariance of $x$ and $y$.
\\
\\
\textbf{Stressed training.} The models might need to generate several consecutive frames in environments of severe packet losses. We can enhance this by adapting a stressed training scheme. We can composite $G$ three times to get $\textbf{y}^{(1)} = G(\textbf{x})$ as before, in addition to $\textbf{y}^{(2)} = G(G(\textbf{x}))$ and $\textbf{y}^{(3)} = G(G(G(\textbf{x})))$. Consequently, we optimise:
\vspace{-0.25cm}
\begin{equation}
\sum_{i=1}^3 \frac{2^{i-1}}{2^3 - 1}  \mathcal{L}(\textbf{x}_{i+1 \cdots T}, \textbf{y}^{(i)}_{1 \cdots T  - i}).
\label{eq:stress}
\end{equation}
This can be generalised for a deeper composition. However, it gets more expensive to train.
\\
\\
\textbf{Data processing.}
The model processes speech with a sliding window of segments of duration 6.25\,ms (corresponding to an array of length 100, in case of a 16\,kHz sample rate). This is close to a typical packet duration of 10-20\,ms \cite{voip, PLC2016}.  This length compromises between two issues. The first is the speed of inference, the second is the number of trainable parameters and generalisation ability. Smaller segment duration needs much longer time for training and inference, because it processes a very long sequence linearly without parallelisation \cite{lstmspeed}. Bigger segments oblige the model to have more parameters, which is more difficult to train. Furthermore, during training, the tracks are segmented into segments of length 20\,s to allow fast training.
\\
\\
\textbf{Hyperparameters.}
The hyperparameters space is explored using BOHB \cite{BOHB}, a state-of-the-art tool for hyperparameter optimisation. The best architecture it discovered is shown in Table \ref{tab:predictive_coding}. The training is performed applying an Adam optimiser \cite{Adam} using the stressed training loss in Equation \ref{eq:stress}. To speed up training, we first train for $80$ epochs without the \textit{stress training}. Then, we continue with the \textit{stress training} for $40$ epochs. For \textit{stress training}, we use a learning rate $\alpha = 0.003$, otherwise $\alpha=0.0045$.
A learning-rate decay $0.0015$ is used, in addition to dropout layers \cite{dropout} of a dropout rate $0.5$ that are entered after each LSTM layer during training, to reduce overfitting.

\subsection{ConcealNet}

\begin{figure*}[t!]
    \centering
    \includegraphics[width=\textwidth]{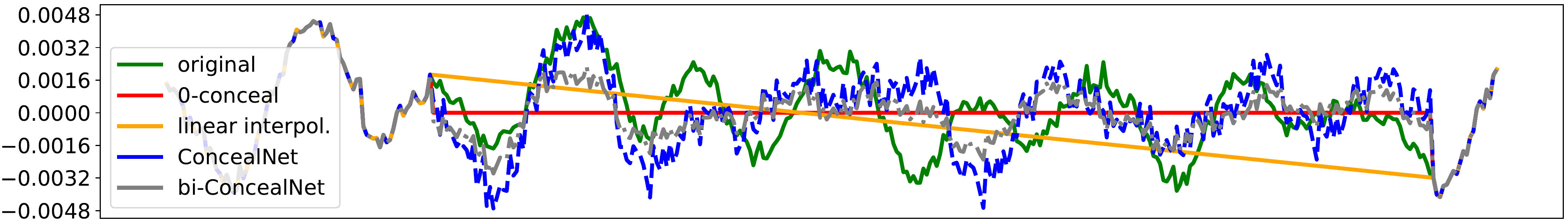}
    \caption{Example of fixed audio segment, demonstrating ConcealNet, bidirectional ConcealNet, the 0-substitution concealment, linear interpolation and original audio signal before loss. This is a segment of length 25ms from the track `valid\_13.wav'.}
    \vspace{-0.2cm}
    \label{fig:example}
\end{figure*}

\subsubsection{Recurrent concealment cells}
\label{subsec:concealment_cell}
Given is an input sequence $\textbf{x}_{1\cdots T}$ with a binary mask $M_{1\cdots T}$, where $M_t = 1$ only if $\textbf{x}_t$ is not lost, otherwise $M_t = 0$ and $\textbf{x}_t = \textbf{0}$. Also, given is a generative recurrent cell $R$ with output value that estimates the next element of the input sequence, namely $\textbf{y}_t = \hat{\textbf{x}}_{t + 1}$. We introduce a wrapper concealment recurrent cell $F_R$ that uses $R$ to fix $\textbf{x}$. The input of $F_R$ is $(\textbf{x}_t, M_t)$, and its previous state $\textbf{z}_{t-1} = (\hat{\textbf{x}}_t, \textbf{s}_{t-1})$. One step of the concealment cell is executed according to the equations:
\begin{align}
\textbf{y}_t &\leftarrow \textbf{x}_t M_t + \hat{\textbf{x}}_{t} (1 - M_t) \\
\hat{\textbf{x}}_{t+1}, \textbf{s}_{t} &\leftarrow R(\textbf{y}_t, \textbf{s}_{t - 1}) \\
\textbf{z}_t &\leftarrow(\hat{\textbf{x}}_{t+1}, \textbf{s}_{t}) \\
F_R(\textbf{x}_t, M_t, \textbf{z}_{t-1}) &\leftarrow (\textbf{y}_t, \textbf{z}_t).
\end{align}

The initial state is given by $\textbf{z}_0 = (\textbf{p}, \textbf{s}_0)$, where $\textbf{p}$ is a default-response vector, in case the initial frames were lost. In our implementation, we use $\textbf{p} = \textbf{0}$. 

The value of $\textbf{y}_t$ will be the same as $\textbf{x}_t$ if $M_t = 1$ (non-lost frame). Otherwise, it will be the generated value $\hat{\textbf{x}}_t$, which is the predicted element according to the cell $R$. After that, the predicted next element $\hat{\textbf{x}}_{t+1}$ is computed using $R$, in addition to the state $\textbf{s}_t$ that will be used in the next time step. A visual demonstration of how this cell operates is depicted in Figure \ref{fig:concealment}.

This cell behaves similar to the PLC algorithm in 
\cite{Lotfidereshgi2018}. However, formulating it as a neural operation allows the models to perform PLC end-to-end from corrupt raw audio to concealed raw audio. This can be embedded to make end-to-end inference, from lossy speech to emotions directly (or other tasks).

\subsubsection{End-to-end PLC inference}
\label{subsec:concealnet} 
Putting together the aforementioned components, we extract the recurrent cells and Fully Connected layers \cite{FC} from the generative RNN $G$ trained in Section \ref{subsec:generativeRNN}. The extracted cells are then stacked to form one cell $R$, and then we wrap it using the concealment wrappers introduced in Subsection \ref{subsec:concealment_cell}. Consequently, we construct an end-to-end PLC inference RNN (which we call ConcealNet) with the input sequence $\textbf{x}_{1\cdots T}$, and a corresponding loss mask $M_{1\cdots T}$ to predict a fixed signal $\hat{\textbf{x}}_{1\cdots T}$. This resulting fixed signal is after applying PLC, where lost segments are concealed and non-lost segments are copied. 
\\
\\
\textbf{Bidirectional concealment.}
Bidirectional RNNs have shown promising improvements in ASR \cite{bilstm}, which motivates us to introduce a bidirectional variant 
assuming non-causal processing is an option, \eg by a small buffer or in post-hoc application.  
If we train a backwards generative network and use the same architecture of ConcealNet on it (by reversing the input and output sequences), the results of those two networks (forward and backward) can be merged by averaging both to obtain a simple bidirectional variant of ConcealNet. This variant tends to have better performance generally. However, its main disadvantage is the inability to be used in real-time settings, because it assumes the knowledge about future context. 

\begin{table}[b!]
\centering

\begin{tabular}{c c c | c c c c}

\multicolumn{3}{c|}{$\mathcal{M}$ parameters} & \multicolumn{4}{c}{audio} \\
\hline
$p_L$ & $p_N$ & $\downarrow$ drop\,\% & 0-conc & interp & Forw  & Bidir  \\
\hline
0.1 & 0.9 & 10.16 & 94.65 & 91.86 & 98.54 & \textbf{98.99} \\
0.5 & 0.9 & 17.16 & 90.58 & 85.53 & 95.16 & \textbf{96.69} \\
0.1 & 0.5 & 35.83 & 78.21 & 69.82 & 94.14 & \textbf{95.76} \\
0.1 & 0.1 & 50.06 & 66.67 & 56.57 & 90.70 & \textbf{92.98} \\
0.5 & 0.5 & 50.28 & 66.41 & 55.09 & 83.97 & \textbf{88.39} \\
0.9 & 0.9 & 50.41 & \textbf{66.07} & 54.32 & 56.15 & 64.69 \\
0.5 & 0.1 & 64.52 & 52.28 & 40.58 & 76.27 & \textbf{81.54} \\
0.9 & 0.5 & 83.58 & 28.29 & 19.27 & 28.25 & \textbf{34.49} \\
0.9 & 0.1 & 90.15 & 17.93 & 11.69 & 18.46 & \textbf{22.55} \\

\end{tabular}
\vspace{0.1cm}
\caption{CCC percentage scores of comparing concealed audio to original audio, after using different strategies for PLC.}
\vspace{-0.5cm}
\label{tab:audio_results}
\end{table}

\subsection{Dataset}
The dataset that is used in the experiment is the RECOLA dataset \cite{RECOLA}. The dataset consists of  16 training tracks and 15 validation tracks. Each track consists of 5 minutes of audio \cite{RECOLA}, we downsampled them to 16\,kHz. Each track is labelled with emotions across time and the labels were collected on a frequency of 25\,Hz. However, we reduced this into 5\,Hz using median pooling. Emotions are represented as two main features, namely \textit{arousal} and \textit{valence}.

\subsection{Emotion model}
For emotions predictions, we use a state-of-the-art end-to-end model \cite{tzirakis2018} to predict emotions from raw audio. 
The model predicts two dimensions across time, namely \textit{arousal} and \textit{valence}. The architecture we use consists of 3 convolution blocks, followed by 2 LSTM layers of 85 units, then a Fully Connected layer of 65 units, and a final output layer. Each convolution block consists of a convolution layer of 47 output channels followed by max pooling. The kernel sizes are $(27, 14, 3)$, and the pooling sizes are $(40, 20, 4)$ for the 3 blocks respectively.

This emotions model is appended to the ConcealNet presented in Subsection \ref{subsec:concealnet} to make end-to-end predictions of emotions from speech with lossy packets.

\subsection{Packet loss generation}

To simulate the behaviour of lossy and non-lossy packets in a given sequence, we adapt the Markov Chain 
$\mathcal{M}(p_{\text{L}}, p_{\text{N}})$ as shown in Figure \ref{fig:markov}. \cite{gilbert} has shown it to be an effective approach for packet loss modelling; other models exist like a three-state model \cite{milner2004analysis} to model burst behaviour. \cite{da2019mac} reviews other models.
Given a sequence of $T$ frames, we sample a binary mask $M$ by starting at the state $N$, then transitioning between the states $N$ (for no-loss) and $L$ (for loss) based on the transition probabilities until $T$ states are enumerated. The sampled sequence of states is directly transformed into the mask $M$.

\subsection{Baselines}

\subsubsection{0-substitution concealment}
This is a simple baseline, which replaces all the loss values by one constant value, which is 0 \cite{perkins1998survey}. Even though this baseline is very simple, it serves the purpose of showing how important it is to solve the concealment problem and how far it can be improved.

\subsubsection{Linear interpolation}

Linear interpolation is a technique which conceals a lost segment using a linear equation joining the last point before the loss and the first point after the loss, and then predicting the lost values in between, according to the equation \cite{interpolation}.

\section{Experiments and Results}
\label{sec:results}

In order to evaluate different methods for Packet Loss Concealment (PLC), first, we use input signals $\textbf{x}$ with a corresponding loss mask $M$ (sampled by the two-state Markov Chain) and corresponding ground truth emotions labels $\textbf{y}$.  Then, we use the end-to-end model to acquire the concealed signal $\hat{\textbf{x}}$ and the corresponding emotions labels $\hat{\textbf{y}}$. Consequently, we compare $\hat{\textbf{x}}$ against $\textbf{x}$ to examine the quality of the concealment, and $\hat{\textbf{y}}$ against $\textbf{y}$ to examine the quality of the emotions prediction after PLC.
In all scenarios, we use CCC \cite{ccc} as the comparison metric, since it measures data reproducibility. The results of the concealment's quality on the audio data are shown in Table \ref{tab:audio_results}, in addition to Table \ref{tab:emo_results}, which shows the results of the emotions predictions after concealment. Eventually, we show the results of the \textit{stress training} scheme in Table \ref{tab:stress_results}.

\begin{table*}[t!]
\centering
\begin{tabular}{c c c | c c c c | c| c c c c | c}
\multicolumn{3}{c|}{$\mathcal{M}$ parameters} & \multicolumn{5}{c|}{Arousal} & \multicolumn{5}{c}{Valence} \\
\hline
$p_L$ & $p_N$ & $\downarrow$ drop\,\% & 0-conc & interp & Forw  & Bidir & orig & 0-conc & interp & Forw  & Bidir & orig \\
\hline
0.1 & 0.9 & 10.16 & 73.83 & 74.06 & 76.66 & \textbf{76.86} & 76.93 & 38.62 & 38.58 & 42.98 & \textbf{43.11} & 43.18 \\
0.5 & 0.9 & 17.16 & 68.41 & 70.36 & 75.99 & \textbf{76.58} & 76.93 & 28.02 & 31.05 & 42.35 & \textbf{42.86} & 43.18 \\
0.1 & 0.5 & 35.83 & 68.09 & 66.59 & 76.18 & \textbf{76.99} & 76.93 & 26.70 & 23.52 & 40.52 & \textbf{41.95} & 43.18 \\
0.1 & 0.1 & 50.06 & 69.15 & 62.34 & 76.55 & \textbf{77.50} & 76.93 & 25.29 & 16.92 & 39.81 & \textbf{41.19} & 43.18 \\
0.5 & 0.5 & 50.28 & 58.26 & 59.32 & 73.42 & \textbf{75.38} & 76.93 & 13.96 & 14.11 & 37.62 & \textbf{39.90} & 43.18 \\
0.9 & 0.9 & 50.41 & 63.08 & 67.13 & \textbf{69.84} & 67.33 & 76.93 & 17.43 & 22.04 & \textbf{36.70} & 35.05 & 43.18 \\
0.5 & 0.1 & 64.52 & 59.04 & 55.47 & 73.43 & \textbf{75.01} & 76.93 & 13.71 & 11.32 & 35.93 & \textbf{36.61} & 43.18 \\
0.9 & 0.5 & 83.58 & 56.36 & 59.75 & \textbf{67.55} & 64.11 & 76.93 & 10.41 & 11.59 & \textbf{26.55} & 22.77 & 43.18 \\
0.9 & 0.1 & 90.15 & 58.27 & 59.59 & \textbf{67.74} & 65.50 & 76.93 & 10.47 & 11.19 & \textbf{20.93} & 17.50 & 43.18 \\

\end{tabular}
\vspace{0.25cm}
\caption{CCC percentage scores for arousal and valence prediction after using different strategies for Packet Loss Concealment. 
}
\vspace{-0.5cm}
\label{tab:emo_results}
\end{table*}

\begin{table}[h!]
\centering
\begin{tabular}{c | c c | c c | c c}

 & \multicolumn{2}{c|}{arousal} &  \multicolumn{2}{c|}{valence} &  \multicolumn{2}{c}{audio} \\
\hline
drop\,\% & - & stress & -  & stress & - & stress  \\
\hline
10.16 & 76.62 & \textbf{76.66} & \textbf{43.03} & 42.98 & \textbf{98.66} & 98.54 \\
17.16 & 75.75 & \textbf{75.99} & 41.89 & \textbf{42.35} & 95.13 & \textbf{95.16} \\
35.83 & 75.78 & \textbf{76.18} & \textbf{40.67} & 40.52 & 94.13 & \textbf{94.14} \\
50.06 & 75.61 & \textbf{76.55} & \textbf{39.90} & 39.81 & 89.91 & \textbf{90.70} \\
50.28 & 71.70 & \textbf{73.42} & 35.37 & \textbf{37.62} & 82.86 & \textbf{83.97} \\
50.41 & 57.69 & \textbf{69.84} & 25.20 & \textbf{36.70} & 49.17 & \textbf{56.15} \\
64.52 & 70.33 & \textbf{73.43} & 30.38 & \textbf{35.93} & 72.31 & \textbf{76.27} \\
83.58 & 52.94 & \textbf{67.55} & 15.69 & \textbf{26.55} & 23.56 & \textbf{28.25} \\
90.15 & 53.89 & \textbf{67.74} & 11.76 & \textbf{20.93} & 15.34 & \textbf{18.46} \\

\end{tabular}
\vspace{0.25cm}
\caption{CCC percentage scores of the effects of the \textit{stress training} on ConcealNet, for the different tasks (audio PLC and corresponding emotions predictions).}
\vspace{-0.75cm}
\label{tab:stress_results}
\end{table}


Both versions of ConcealNet are performing much better in all scenarios of emotion prediction. Especially, where $p_L$ is not high ($leq 0.5$), both versions of ConcealNet have a small drop in emotions predictions, even when the overall frame drop-rate is up to 64\,\%. For audio concealment, the bidirectional ConcealNet has the best performance, followed by forward ConcealNet. The results are degraded, however, in one scenario when $p_L=0.9$, the 0-concealment baseline achieves the best results in the loss concealment. 

The scenario where $p_L$ is high is the scenario where ConcealNet experiences relatively long loss, and it is expected to recover $p_L / (1 - p_L) \sim 9$ consecutive segments, for each loss occurrence, which is extremely challenging. However, we observe how the \textit{stress training} has managed to conquer this problem, as shown by the improvements in Table \ref{tab:stress_results}, where the stress trained models are generally overperforming especially in the scenarios with more losses.


\section{Conclusions}
\label{sec:conclusion}
In this paper, a concealment RNN (ConcealNet) was introduced. This consists of two main components: the first is a stacked generative recurrent cell $R$ which is trained to predict elements of sequences given the preceding elements, and a wrapper $F_R$ for such a stacked cell. The wrapped recurrent cell can be used as a recurrent layer given an input sequence $\textbf{x}$ and a corresponding binary mask $M$ marking losses, to output a concealed sequence $\hat{\textbf{x}}$. A generative RNN consisting of two LSTM layers was trained to be used by ConcealNet, in addition to an emotions model which was connected to the ConcealNet, to conceal audio and predict emotions end-to-end. A stress training scheme was introduced to improve the performance of ConcealNet on long-term losses. Furthermore, the proposed ConcealNet was used in two variants, one processing the sequence forwards and the other processing the sequence bidirectionally by averaging forwards and backwards.
The fully reproducible experiments on the popular RECOLA continuous emotion database have shown that the proposed ConcealNet is getting considerably good results in scenarios without too long losses, even when they are frequent. In environments with short packet losses, after using ConcealNet, the degradation of speech emotion prediction is minor: for arousal, CCC dropped from 76.93\,\% to 75.99\,\%, while for valence, it dropped from 43.18\,\% to 39.81\,\%. The bidirectional variant of ConcealNet is achieving even better results. The scenario when there are long packet losses has been shown to be the most challenging as one may expect. However, a stress training technique was introduced to conquer this issue and it has shown an improvement of the results.

Future work can consider the usage of attention mechanisms and the introduction of generative approaches such as variants of generative adversarial topologies or variational solutions.



\bibliographystyle{IEEEtran}
\bibliography{ms}

\begin{thebibliography}{10}
\providecommand{\url}[1]{#1}
\csname url@samestyle\endcsname
\providecommand{\newblock}{\relax}
\providecommand{\bibinfo}[2]{#2}
\providecommand{\BIBentrySTDinterwordspacing}{\spaceskip=0pt\relax}
\providecommand{\BIBentryALTinterwordstretchfactor}{4}
\providecommand{\BIBentryALTinterwordspacing}{\spaceskip=\fontdimen2\font plus
\BIBentryALTinterwordstretchfactor\fontdimen3\font minus
  \fontdimen4\font\relax}
\providecommand{\BIBforeignlanguage}[2]{{%
\expandafter\ifx\csname l@#1\endcsname\relax
\typeout{** WARNING: IEEEtran.bst: No hyphenation pattern has been}%
\typeout{** loaded for the language `#1'. Using the pattern for}%
\typeout{** the default language instead.}%
\else
\language=\csname l@#1\endcsname
\fi
#2}}
\providecommand{\BIBdecl}{\relax}
\BIBdecl

\bibitem{voip}
A.~Takahashi, H.~Yoshino, and N.~Kitawaki, ``{Perceptual QoS assessment
  technologies for VoIP},'' \emph{IEEE Communications Magazine}, vol.~42,
  no.~7, pp. 28--34, 2004.

\bibitem{hmminterspeech}
B.~J. Borgstr{\"o}m, P.~H. Borgstr{\"o}m, and A.~Alwan, ``{Efficient HMM-Based
  Estimation of Missing Features, with Applications to Packet Loss
  Concealment},'' in \emph{{Proceedings INTERSPEECH, 11th Annual Conference of
  the International Speech Communication Association}}.\hskip 1em plus 0.5em
  minus 0.4em\relax Makuhari, Chiba, Japan: ISCA, 2010, pp. 2394--2397.

\bibitem{schuller2013}
B.~W. Schuller, \emph{{Intelligent Audio Analysis}}.\hskip 1em plus 0.5em minus
  0.4em\relax Springer Publishing Company, Incorporated, 2013.

\bibitem{codec}
A.~Janicki and B.~Ksiundefinedundefinedak, ``{Packet Loss Concealment Algorithm
  for VoIP Transmission in Unreliable Networks},'' in \emph{{Proceedings of the
  2008 Conference on New Trends in Multimedia and Network Information
  Systems}}.\hskip 1em plus 0.5em minus 0.4em\relax NLD: IOS Press, 2008, pp.
  23–--33.

\bibitem{graves}
A.~Graves, ``{Generating Sequences With Recurrent Neural Networks},''
  \emph{CoRR}, vol. abs/1308.0850, 2013.

\bibitem{GANS}
I.~Goodfellow, J.~Pouget-Abadie, M.~Mirza, B.~Xu, D.~Warde-Farley, S.~Ozair,
  A.~Courville, and Y.~Bengio, ``{Generative Adversarial Nets},'' in
  \emph{{Advances in Neural Information Processing Systems 27}}, Z.~Ghahramani,
  M.~Welling, C.~Cortes, N.~D. Lawrence, and K.~Q. Weinberger, Eds.\hskip 1em
  plus 0.5em minus 0.4em\relax Montreal, Canada: Curran Associates, Inc., 2014,
  pp. 2672--2680.

\bibitem{Lotfidereshgi2018}
R.~Lotfidereshgi and P.~Gournay, ``{Speech Prediction Using an Adaptive
  Recurrent Neural Network with Application to Packet Loss Concealment},'' in
  \emph{{IEEE International Conference on Acoustics, Speech and Signal
  Processing (ICASSP)}}.\hskip 1em plus 0.5em minus 0.4em\relax Calgary, AB,
  Canada: IEEE, 2018, pp. 5394--5398.

\bibitem{mywork}
M.~M. Mohamed and B.~W. Schuller, ``{I have vxxx bxx connexxxn!: Facing Packet
  Loss in Deep Speech Emotion Recognition},'' \emph{arXiv preprint arXiv},
  2020.

\bibitem{Marchi16-RTO}
E.~Marchi, F.~Eyben, G.~Hagerer, and B.~W. Schuller, ``{Real-time Tracking of
  Speakers' Emotions, States, and Traits on Mobile Platforms},'' in
  \emph{{Proceedings INTERSPEECH, 17th Annual Conference of the International
  Speech Communication Association}}.\hskip 1em plus 0.5em minus 0.4em\relax
  San Francisco, CA: ISCA, 2016, pp. 1182--1183.

\bibitem{smartphones}
A.~Ignatov, R.~Timofte, A.~Kulik, S.~Yang, K.~Wang, F.~Baum, M.~Wu, L.~Xu, and
  L.~Van~Gool, ``{AI Benchmark: All About Deep Learning on Smartphones in
  2019},'' \emph{arXiv preprint arXiv:1910.06663}, 2019.

\bibitem{PLC2016}
B.-K. Lee and J.-H. Chang, ``{Packet Loss Concealment Based on Deep Neural
  Networks for Digital Speech Transmission},'' \emph{IEEE/ACM Trans. Audio,
  Speech and Lang. Proc.}, vol.~24, no.~2, p. 378–387, 2016.

\bibitem{TIMIT}
J.~S. Garofolo, L.~F. Lamel, W.~M. Fisher, J.~G. Fiscus, and D.~S. Pallett,
  ``Darpa timit acoustic-phonetic continous speech corpus cd-rom. nist speech
  disc 1-1.1,'' \emph{NASA STI/Recon technical report n}, vol.~93, 1993.

\bibitem{GAN}
Y.~{Shi}, N.~{Zheng}, Y.~{Kang}, and W.~{Rong}, ``{Speech Loss Compensation by
  Generative Adversarial Networks},'' in \emph{{Asia-Pacific Signal and
  Information Processing Association Annual Summit and Conference (APSIPA
  ASC)}}.\hskip 1em plus 0.5em minus 0.4em\relax Lanzhou, China: IEEE, 2019,
  pp. 347--351.

\bibitem{acousticinpainting}
P.~P. Ebner and A.~Eltelt, ``{Audio inpainting with generative adversarial
  network},'' \emph{arXiv preprint arXiv:2003.07704}, 2020.

\bibitem{xiao2018packet}
X.~Xiao and S.~Zarar, ``{Packet loss concealment with recurrent neural networks
  for wireless inertial pose tracking},'' in \emph{{Proceedings IEEE 15th
  International Conference on Wearable and Implantable Body Sensor Networks
  (BSN)}}.\hskip 1em plus 0.5em minus 0.4em\relax Las Vegas, NV, USA: IEEE,
  2018, pp. 25--29.

\bibitem{khan2018}
N.~M. Khan and G.~M. Khan, ``{Signal Reconstruction Using Evolvable Recurrent
  Neural Networks},'' in \emph{{Intelligent Data Engineering and Automated
  Learning -- IDEAL 2018}}, H.~Yin, D.~Camacho, P.~Novais, and A.~J.
  Tall{\'o}n-Ballesteros, Eds.\hskip 1em plus 0.5em minus 0.4em\relax Cham:
  Springer International Publishing, 2018, pp. 594--602.

\bibitem{Miller2011}
J.~F. Miller, ``Cartesian genetic programming,'' in \emph{Cartesian Genetic
  Programming}.\hskip 1em plus 0.5em minus 0.4em\relax Springer, 2011, pp.
  17--34.

\bibitem{mack2019deep}
W.~Mack and E.~A. Habets, ``{Deep filtering: Signal extraction using complex
  time-frequency filters},'' \emph{arXiv preprint arXiv:1904.08369}, p. CoRR,
  2019.

\bibitem{deeplearn}
I.~J. Goodfellow, Y.~Bengio, and A.~Courville, \emph{{Deep Learning}}.\hskip
  1em plus 0.5em minus 0.4em\relax Cambridge, MA, USA: MIT Press, 2016.

\bibitem{LSTM}
S.~Hochreiter and J.~Schmidhuber, ``{Long Short-term Memory},'' \emph{Neural
  computation}, vol.~9, no.~8, pp. 1735--1780, 1997.

\bibitem{GRU}
K.~Cho, B.~van Merrienboer, C.~Gulcehre, D.~Bahdanau, F.~Bougares, H.~Schwenk,
  and Y.~Bengio, ``{Learning Phrase Representations using RNN Encoder-Decoder
  for Statistical Machine Translation},'' 2014.

\bibitem{ccc}
I.~Lawrence and K.~Lin, ``{A concordance correlation coefficient to evaluate
  reproducibility},'' \emph{Biometrics}, pp. 255--268, 1989.

\bibitem{lstmspeed}
K.~Hwang and W.~Sung, ``{Single stream parallelization of generalized LSTM-like
  RNNs on a GPU},'' in \emph{{IEEE International Conference on Acoustics,
  Speech and Signal Processing (ICASSP)}}.\hskip 1em plus 0.5em minus
  0.4em\relax Brisbane, QLD, Australia: IEEE, 2015, pp. 1047--1051.

\bibitem{BOHB}
S.~Falkner, A.~Klein, and F.~Hutter, ``{BOHB:} robust and efficient
  hyperparameter optimization at scale,'' \emph{CoRR}, vol. abs/1807.01774,
  2018.

\bibitem{Adam}
D.~P. Kingma and J.~Ba, ``{Adam: A Method for Stochastic Optimization},'' in
  \emph{{3rd International Conference on Learning Representations, Conference
  Track Proceedings}}, Y.~Bengio and Y.~LeCun, Eds.\hskip 1em plus 0.5em minus
  0.4em\relax San Diego, CA, USA: ICLR, 2015.

\bibitem{dropout}
N.~Srivastava, G.~Hinton, A.~Krizhevsky, I.~Sutskever, and R.~Salakhutdinov,
  ``{Dropout: a simple way to prevent neural networks from overfitting},''
  \emph{The journal of machine learning research}, vol.~15, no.~1, pp.
  1929--1958, 2014.

\bibitem{FC}
D.~E. Rumelhart, G.~E. Hinton, and R.~J. Williams, ``{Learning internal
  representations by error propagation},'' California Univ San Diego La Jolla
  Inst for Cognitive Science, Tech. Rep., 1985.

\bibitem{bilstm}
A.~{Zeyer}, P.~{Doetsch}, P.~{Voigtlaender}, R.~{Schlüter}, and H.~{Ney}, ``{A
  comprehensive study of deep bidirectional LSTM RNNS for acoustic modeling in
  speech recognition},'' in \emph{{Proceedings IEEE International Conference on
  Acoustics, Speech and Signal Processing (ICASSP)}}.\hskip 1em plus 0.5em
  minus 0.4em\relax New Orleans, LA, USA: IEEE, 2017, pp. 2462--2466.

\bibitem{RECOLA}
F.~Ringeval, A.~Sonderegger, J.~S. Sauer, and D.~Lalanne, ``{Introducing the
  RECOLA multimodal corpus of remote collaborative and affective
  interactions},'' \emph{2013 10th IEEE International Conference and Workshops
  on Automatic Face and Gesture Recognition (FG)}, pp. 1--8, 2013.

\bibitem{tzirakis2018}
P.~{Tzirakis}, J.~{Zhang}, and B.~W. {Schuller}, ``{End-to-End Speech Emotion
  Recognition Using Deep Neural Networks},'' in \emph{{Proceedings IEEE
  International Conference on Acoustics, Speech and Signal Processing
  (ICASSP)}}.\hskip 1em plus 0.5em minus 0.4em\relax Calgary, AB, Canada: IEEE,
  2018, pp. 5089--5093.

\bibitem{gilbert}
G.~Ha{\ss}linger and O.~Hohlfeld, ``{The Gilbert-Elliott Model for Packet Loss
  in Real Time Services on the Internet},'' in \emph{{14th GI/ITG
  Conference-Measurement, Modelling and Evaluation of Computer and
  Communication Systems}}.\hskip 1em plus 0.5em minus 0.4em\relax Dortmund,
  Germany: VDE, 2008, pp. 1--15.

\bibitem{milner2004analysis}
B.~P. Milner and A.~B. James, ``{An Analysis of Packet Loss Models for
  Distributed Speech Recognition},'' in \emph{{Proceedings INTERSPEECH, 8th
  International Conference on Spoken Language Processing}}.\hskip 1em plus
  0.5em minus 0.4em\relax Jeju Island, Korea: ISCA, 2004, pp. 1549--1552.

\bibitem{da2019mac}
C.~A.~G. Da~Silva and C.~M. Pedroso, ``{MAC-Layer Packet Loss Models for Wi-Fi
  Networks: A Survey},'' \emph{IEEE Access}, vol.~7, pp. 180\,512--180\,531,
  2019.

\bibitem{perkins1998survey}
C.~Perkins, O.~Hodson, and V.~Hardman, ``{A survey of packet loss recovery
  techniques for streaming audio},'' \emph{IEEE network}, vol.~12, no.~5, pp.
  40--48, 1998.

\bibitem{interpolation}
A.~M. Bayen and T.~Siauw, ``{Chapter 14 - Interpolation},'' in \emph{{An
  Introduction to MATLAB® Programming and Numerical Methods for Engineers}},
  A.~M. Bayen and T.~Siauw, Eds.\hskip 1em plus 0.5em minus 0.4em\relax Boston:
  Academic Press, 2015, pp. 211--223.

\end{thebibliography}




\clearpage

\end{document}